# Superconductor Dynamics

*F. Gömöry*[1]

Institute of Electrical Engineering, Slovak Academy of Sciences, Bratislava, Slovakia

**Abstract**
Superconductors used in magnet technology could carry extreme currents because of their ability to keep the magnetic flux motionless. The dynamics of the magnetic flux interaction with superconductors is controlled by this property. The cases of electrical transport in a round wire and the magnetization of wires of various shapes (circular, elliptical, plate) in an external magnetic field are analysed. Resistance to the magnetic field penetration means that the field produced by the superconducting magnet is no longer proportional to the supplied current. It also leads to a dissipation of electromagnetic energy. In conductors with unequal transverse dimensions, such as flat cables, the orientation with respect to the magnetic field plays an essential role. A reduction of magnetization currents can be achieved by splitting the core of a superconducting wire into fine filaments; however, new kinds of electrical currents that couple the filaments consequently appear. Basic formulas allowing qualitative analyses of various flux dynamic cases are presented.

## 1   Introduction

The main motivation for using type II superconductors (in these materials the magnetic flux is divided into flux quanta called flux lines) in electromagnets originates in their ability to carry large electrical currents without dissipation. This property, which enables the generation of stationary magnetic fields beyond the reach of traditional technology based on metallic conductors, is the consequence of a mechanism that hinders the motion of magnetic flux inside such a superconductor [1, 2]. Flux lines interact with pinning centres that usually are imperfections in the composition or in the structure. This phenomenon is illustrated schematically in Fig. 1, which shows a superconducting plate with finite thickness in the *x*-direction and very large dimensions along the other two axes exposed to a magnetic field parallel to the *z*–*y* plane. For simplicity, just one kind of pinning centre is shown on the right-hand side of Fig. 1 (see the white objects of irregular shape randomly distributed in the volume of the superconductor). When the applied magnetic field, $B_{ex}$, is switched on, the magnetic flux enters the plate in the form of flux lines, each carrying the elementary quantum of magnetic flux, $\Phi_0 = 2 \times 10^{-15}$ Wb. A rough estimate of the distance between the flux lines can be obtained by assuming the absence of any mechanism hindering the flux penetration, i.e. neglecting also a flux pinning. In that case, the lines would distribute uniformly and the flux density inside the plate would be roughly equal to the external applied field, $B_{ex}$. Let us assume that the flux lines form the quadratic lattice shown in Fig. 2. Its line spacing could be then calculated as

$$a = \sqrt{\frac{\Phi_0}{B_{ex}}}, \tag{1}$$

---

[1] elekgomo@savba.sk

resulting in the applied field $B_{ex}$ = 1 T at distance $a$ = 45 nm. In other words, there are a very large number of flux lines interacting with a superconducting wire with a typical diameter of 1 mm. This conclusion remains valid after detailed calculations, indicating that a hexagonal lattice would be preferred. Now consider the influence of flux pinning. Because the pinning centres obstruct the flux penetration, we will find the highest density of flux lines close to the sample surface, and it will decrease inwards. By applying the Maxwell equation in quasi-static conditions (i.e. neglecting the displacement current) and taking into account all the symmetries, we find for the plate shown in Fig. 1 that

$$\frac{dB_y(x)}{dx} = -j_z(x). \qquad (2)$$

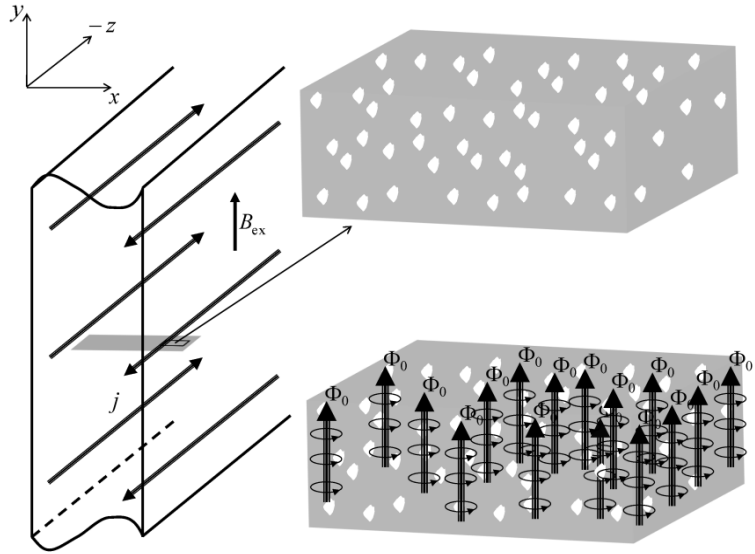

**Fig. 1:** Left: slab from a hard superconductor exposed to a parallel magnetic field $B_{ex}$. Right: the existence of imperfections (white spots in the upper figure) causes a non-uniformity in the concentration of magnetic flux lines (lower figure). Magnetic flux cannot move freely inside a hard superconductor because the imperfections act as pinning centres. Macroscopic currents resulting from the gradient magnetic flux density are indicated in the left-hand picture.

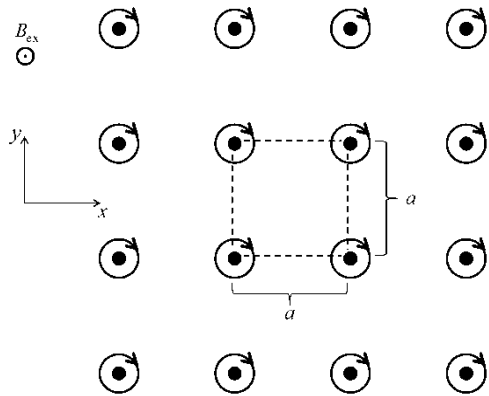

**Fig. 2:** Regular lattice of flux lines created in a type II superconductor when exposed to the magnetic field in the $z$-direction. The central parts (black) denote the normal cores surrounded by circulating supercurrents.

This expression can be interpreted in the following way: a macroscopic electrical current is equivalent to a gradient in the magnetic flux density. It is valid for both normal conductors and superconductors. In a metal the currents (called the eddy currents) dissipate because of normal metal resistivity and the field gradients disappear with time. Only in AC fields with periods shorter than the dissipation time will field gradients persist, leading to the so-called 'skin effect'. However, in the superconductor shown in Fig. 1 the gradient of the density of flux lines—thus of the flux density itself—persists in the quasi-static conditions of switching on the external field. Thus Eq. (2) predicts the existence of persistent macroscopic currents flowing in the $z$-direction. The density of the flux lines is lower than that equivalent to the applied field. In other words, the magnetic field in a superconducting plate is 'screened' by these currents, which produce a magnetic field with a polarity that opposes the applied field $B_{ex}$.

This simple example shows that the 'freezing' of the flux line positions by pinning forces is essential to reach the large stationary densities of electrical current necessary in the construction of electromagnets able to generate static magnetic fields exceeding 10 T. However, the mechanism that serves to keep the distribution of the magnetic field and current density unchanged opposes any change in the magnetic field. Such superconductors are referred to as 'hard' because in a cyclic magnetization process they exhibit hysteresis resembling the behaviour of a magnetic material with large coercivity. The mechanism of flux pinning, however, causes hysteresis of the magnetic flux distribution with respect to the current supplied to the magnet winding. As a consequence, the dynamics of the magnetic flux movement in an electromagnet made from a hard superconductor is controlled by the flux pinning mechanism.

## 2    Critical state model

The interaction of flux lines with pinning centres is understood for some simple cases, and there is extensive research activity dedicated to this issue. Nevertheless, the complexity of superconducting materials and the variety of pinning centres encountered make it difficult to establish a procedure allowing the prediction of the macroscopic volume density of the pinning force directly from microscopic calculations. It is more practical to formulate a phenomenological model describing the macroscopic behaviour of a hard superconductor. Its simplest version was developed by Bean [3], and is commonly known as the 'critical state model'. It is valid on a macroscopic scale, ignoring the details of a magnetic field distribution around an individual flux line. Working with an average taken over many flux lines, the critical state model states that in any (macroscopic) part of a hard superconductor one can find either no electrical current or a current with density equal to the so-called 'critical current density', $j_c$. In the original formulation its value is constant and it characterizes fully the properties of a material that is a hard superconductor, i.e. a type II superconductor with flux pinning. Several later adjustments were incorporated in order to include the properties of superconductors in a more realistic way, e.g. the suppression of the critical current density by the magnetic field [4] or its dependence on the electrical field [5, 6]. With these modifications the results can be better compared to experimental data; however, the results are no longer analytical. Therefore the original version with $j_c$ = const. is still used to obtain a basic understanding of the problem, and we will use this in the following.

The actual value of the current density in a hard superconductor is controlled by the preceding history. No current flows in the regions where the magnetic flux vortices had not penetrated before, and in the rest of the superconductor the flux line density gradient provides a macroscopic current density according to Eq. (2). Because the density of the flux lines controls the magnetic field in a superconductor, the critical state model states that

$$|j| = \begin{cases} 0 & \text{in the places where } B = 0, \\ j_c & \text{elsewhere.} \end{cases} \quad (3)$$

Interestingly, the electrical field that is commonly used to calculate electrical currents in normal metals is not explicitly found in this expression. However, because another Maxwell equation states that any change in magnetic field is accompanied by the appearance of an electrical field, all the regions in a hard superconductor in which a non-zero current density appears will have had to experience a finite electrical field during the transient process of magnetic field penetration. In fact, this is the way to include the history of the magnetization process into the analysis of flux dynamics in hard superconductors.

In spite of its simplicity, the critical state model defined by Eq. (3) has provided many useful predictions that can also be reached in an analytical way.

## 2.1 Magnetic flux dynamics at current transport

Let us consider a long wire with circular cross-section (its length, $l$, significantly exceeds its radius, $R$), made from a hard superconductor characterized by the critical current density $j_c$. The maximum value of DC current that the wire can carry without resistance is

$$I_c = j_c \pi R^2 . \tag{4}$$

When the current $I_1$ (lower than $I_c$) is supplied to the wire in the stand-alone conditions (no additional magnetic field applied), only a part of its cross-section will carry the electrical current and experience a non-zero magnetic field. The process of injecting $I_1$ (through the terminations that are at the wire ends, far from the portion of wire under investigation) is accompanied by the appearance of a temporary electrical field that provokes electrical currents that could not be deleted because of flux pinning. The only solution compatible with the critical state equation (2) is the one shown in the middle part of Fig. 3: a current with density $j_c$ fills the outer shell with a cross-section that is just right to collect the necessary value of transported current. The existence of parallel paths along the wire axis with zero current density (in the centre) and non-zero current density (in the outer shell) does not violate any law of electromagnetism, because in the stationary state the electrical field is zero in the current-free central zone as well as in the outer shell—the latter is a consequence of superconductor zero resistivity. Nevertheless, in the transient process of establishing the transport current, there has been an electrical field in the outer shell. Calculating its distribution in space and time together with the time evolution of the current density distribution allows the determination of the cost of electromagnetic energy (furnished by the supply of wire current) necessary to establish the final distribution [7]. A further increase in current to the value $I_2 > I_1$ (but still below $I_c$) will produce distributions similar to that shown in the centre of Fig. 3, with the subsequent shrinking of the central current-free zone. Eventually, when the transport current reaches $I_c$, all the wire section will be filled by the critical current density, $j_c$.

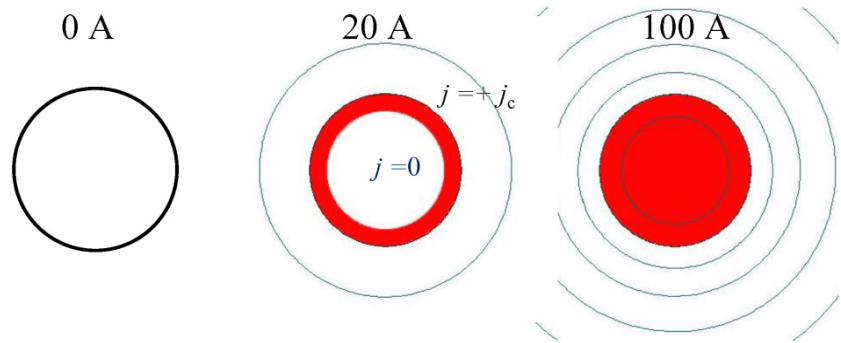

**Fig. 3:** Series of distributions of magnetic field (lines of constant vector potential) and electrical current density (grey scale) calculated for a round wire of radius $R = 0.5$ mm and a critical current of $I_c = 100$ A during the first increase of transported current.

Probably the most striking consequence of the critical state model is reached when the wire current decreases. Because the critical current density once induced in a hard superconductor cannot be cancelled, the only way to reduce the total current transported in the wire is by reversing the current polarity in a part of it. Similar to the case of current increase, it is the outer shell where this process starts, as can be found through a detailed analysis of the electrical field at the transition of total current in the wire from $I_c$ to $I_3 < I_c$. The series of distributions at decreasing current is shown in Fig. 4. Note that the entire section of wire is filled by the critical current density even when the total current is reduced to zero. The difference between this distribution and the original one shown on the left of Fig. 3 is due to flux pinning. History dependence (hysteresis) in the current distribution indicates that, from the thermodynamic point of view, the process is irreversible and that a part of the electromagnetic energy provided by the current supply has been converted to heat. This quantity is called the AC loss because it is commonly determined in a cyclic process, in this example at the transport of alternating current with sinusoidal waveform.

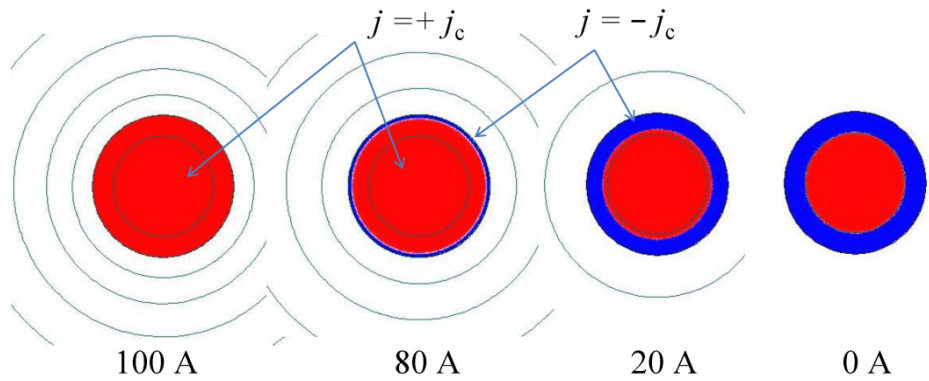

**Fig. 4:** Series of distributions of magnetic field (lines of constant vector potential) and electrical current density (grey scale) calculated for a round wire of radius $R$ = 0.5 mm and a critical current of $I_c$ = 10 A during the reduction of transported current from $I_c$ down to zero.

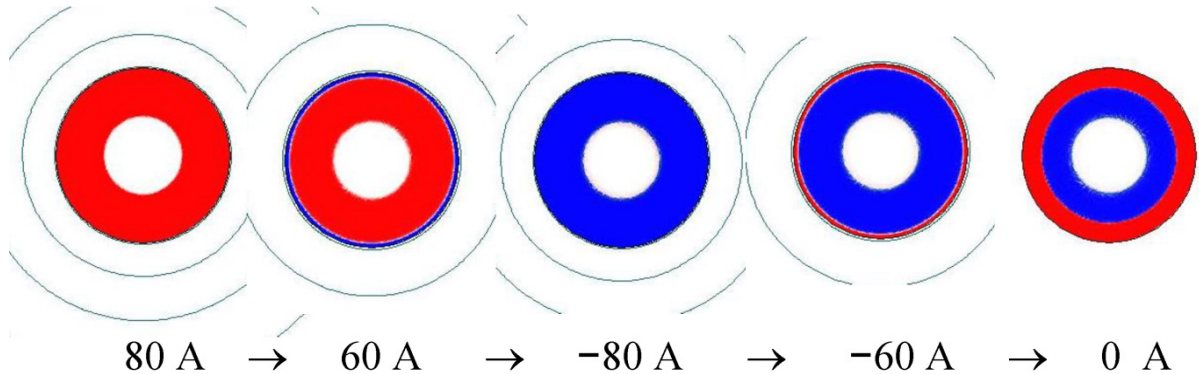

**Fig. 5:** Series of distributions of magnetic field (lines of constant vector potential) and electrical current density (grey scale) calculated for a round wire of radius $R$ = 0.5 mm and a critical current of $I_c$ = 100 A during the sweep of transported current from 80% of $I_c$, to −80%, and then to zero. The central parts (white) denote the current- and field-free neutral zone where the electrical field remained zero throughout.

Usually a superconducting wire operates at transport currents that are always lower than the critical current. Then the central part of the wire remains free of current at all times. Considerations within the critical state model state that no magnetic field (no flux lines) has penetrated to this so-called 'neutral zone', and the electrical field remained zero also during the change of field and current distribution. This property can be used to evaluate the AC loss in the wire carrying an AC current $I_{ac}(t) = I_a \sin\omega t$ with the amplitude $I_a$ and frequency $f = \omega/2\pi$. In fact, the voltage measured on such a wire can be derived from the path integral of the electrical field on the rectangular loop shown in Fig. 6. Because the inner leg of the loop is in the neutral zone, the electric field on that part is zero. Two lines perpendicular to the wire axis do not contribute either, because the electrical field in the axial direction could not exist in this geometry. All the voltage induced in the loop due to the change of the linked magnetic flux with time is on the outer surface of the wire and can be picked by two voltage taps:

$$U = -\frac{\partial \Phi}{\partial t}. \tag{5}$$

This relation is very useful in expressing the loss of electromagnetic energy during one cycle with period $T = 2\pi/\omega$, with the result

$$W = \int_T U I_{ac} \, dt = \int I_{ac} \, d\Phi, \tag{6}$$

which is simply the area of loop enclosed by the $\Phi(I_{ac})$ dependence during one cycle of AC current. The loops for two cycles are shown in Fig. 7. The absence of time as an independent variable means that the shape of this loop does not depend on the frequency $f = \omega/2\pi$. This feature is common for all cases for which the flux pinning controls the flux penetration. The accompanying dissipation is therefore called the hysteresis loss.

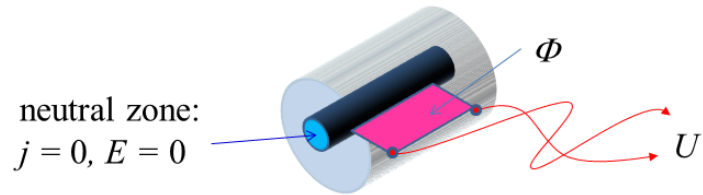

**Fig. 6:** The voltage measured by a pair of taps on the surface of a round wire formed of hard superconductor during transport of electrical current. The voltage can be calculated from the change in magnetic flux enclosed in the rectangular loop consisting of two radial sections perpendicular to the wire axis, with the longitudinal section enclosed in the neutral zone and the line connecting two voltage taps. Electrical fields on all parts except the wire surface are zero.

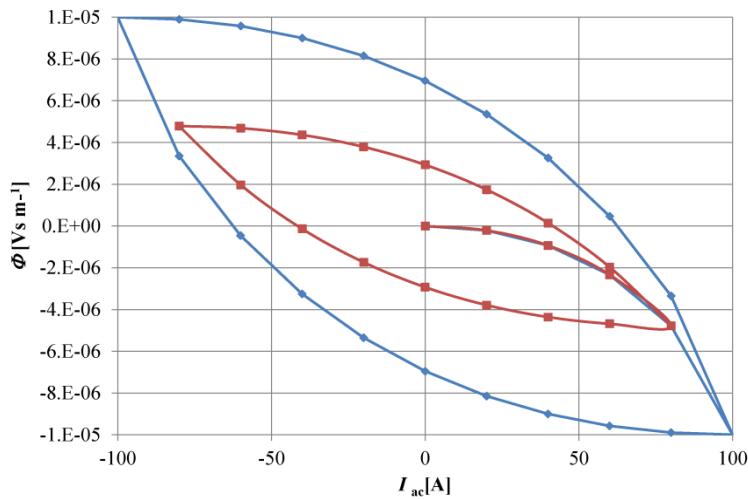

**Fig. 7:** Dependence of magnetic flux enclosed between the wire surface and the neutral zone, $\Phi$, on the wire current, $I_{ac}$, at an amplitude equal to 80% of $I_c$ (squares) and at $I_c$ (diamonds). It begins at (0,0) with the initial part shown in Fig. 3; after the first turn in the current ramping, repetitive cycles with closed loops follow.

## 2.2 Magnetic flux dynamics in transient magnetic field

The example of AC transport by a round wire has demonstrated the main features of magnetic flux dynamics in a hard superconductor. Closer to the actual conditions in the winding of an electromagnet is the case when such a wire is exposed to an external magnetic field that changes in time. In contrast to the previous case, the solutions cannot be found analytically. Fortunately, numerical methods incorporating the critical state principles are available [8–10].

### 2.2.1 Wire with circular cross-section

Figure 8 presents the results of calculations for some significant points in the magnetization cycle of a round wire exposed to a transversal magnetic field. Braking of the magnetic flux penetration is equivalent to the appearance of a current loop (two currents of opposite direction) generating a magnetic field that opposes the change in field inside the superconductor. At the first field increase the central part remains free of currents. This would persist in cases when the maximum field (the amplitude of the field in the case of cyclic AC magnetization) does not exceed the so-called 'penetration field', $B_p$, defined as the field at which all sections of the wire are filled with critical current density.

In order to characterize the process of magnetic flux penetration into the wire exposed to an external magnetic field, the dependence of its magnetization, $M$, on the actual value of the external field, $B_{ex}$, is commonly used. This is illustrated in Fig. 9, where the loop calculated for the wire shown in Fig. 8 is presented. At the penetration field, $B_p$, the sample magnetization reaches the saturation magnetization, $M_s$. Its value can be obtained from the following consideration, illustrated by Fig. 10, which shows the current distribution in a wire with its length parallel to the $z$-coordinate exposed to the magnetic field in the $y$-direction [11]. The magnetic moment will have only the $y$-component that is at any instant of the magnetization cycle calculated by performing the integration (see also Appendix A)

$$m = \int_S -j(x,y)\,dx\,dy \tag{7}$$

over the wire cross-section, $S$. When the saturation state has been reached—this is the situation actually shown in Fig. 10—this expression can be rewritten as

$$m_s = \int_{x<0} -x j_c\,dx\,dy + \int_{x>0} x j_c\,dx\,dy = 2 j_c \int_{x>0} x\,dx\,dy = j_c S \overline{x}, \tag{8}$$

where $\overline{x}$ is the average distance of the $x$-coordinate from the axis of symmetry (in Fig. 10 this is the $y$-axis) in the wire cross-section, i.e.

$$\overline{x} = \frac{1}{S}\int_S |x|\,dS. \tag{9}$$

The quantity $2\overline{x}$ may be interpreted as the wire dimension transverse to the applied field. It can be evaluated in a similar way as for a wire of any shape $S$, although the integration could be more complicated when the boundary between two orientations of current density does not coincide with any of the principal coordinate axes. The saturation magnetization is given by

$$M_s = \frac{m_s}{S} = j_c \overline{x}. \tag{10}$$

For a round wire with radius $R$, one obtains $\overline{x} = \dfrac{4R}{3\pi}$ and $M_s = \dfrac{4R}{3\pi} j_c$, respectively.

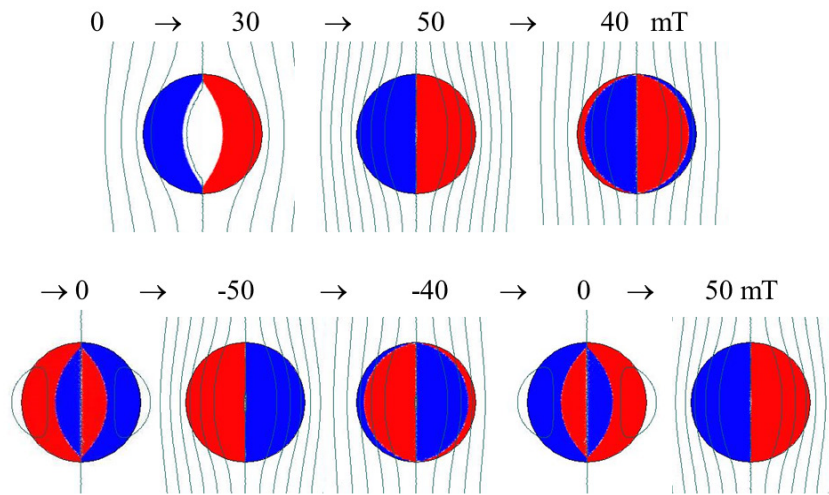

**Fig. 8:** Series of distributions of magnetic field (lines of constant vector potential) and electrical current density (grey scale) calculated for a round wire of radius $R = 0.5$ mm and critical current of $I_c = 100$ A during the application of an external magnetic field, $B_{ex}$, in a direction transverse to the wire axis. Actual values of $B_{ex}$ (in milli-tesla) are indicated above the plots.

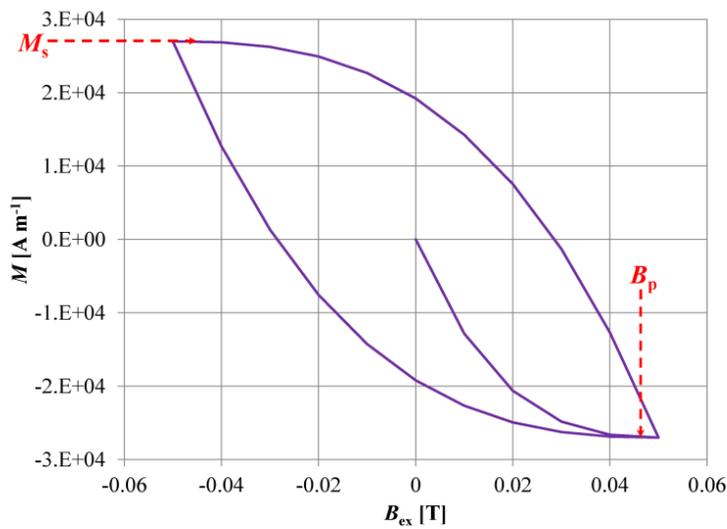

**Fig. 9:** Dependence of magnetization on the applied field calculated for the round wire of Fig. 8

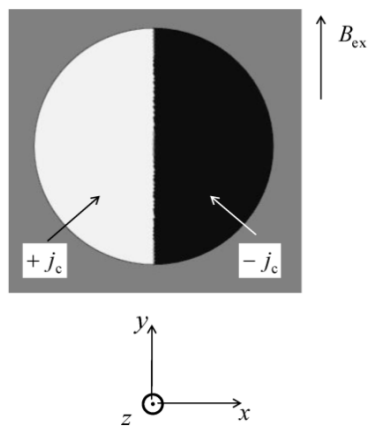

**Fig. 10:** Saturation of round wire with screening currents

### 2.2.2  Wire with elliptical cross-section—effect of magnetic field orientation

The importance of the transverse dimension can be nicely illustrated by the following example. Assume a wire with an elliptical cross-section, the relation between the main axes (commonly called the aspect ratio) being $a:b = 10:1$. Two typical cases of its orientation with respect to the applied field are shown in Fig. 11. When the field is parallel to the minor axis—this is called the 'perpendicular' orientation—the values of the transverse dimension and the saturation magnetization will be

$$\bar{x}_\perp = \frac{4a}{3\pi}; \quad M_s = j_c \frac{4a}{3\pi}, \tag{11}$$

respectively. Rotating the wire by 90° will lead to the configuration called 'parallel', with the following values of transverse dimension and saturation magnetization:

$$\bar{x}_\parallel = \frac{4b}{3\pi}; \quad M_s = j_c \frac{4b}{3\pi}. \tag{12}$$

Because $a > b$, the values for the perpendicular field are larger than those for the parallel field, by exactly the same ratio as the ellipse axes.

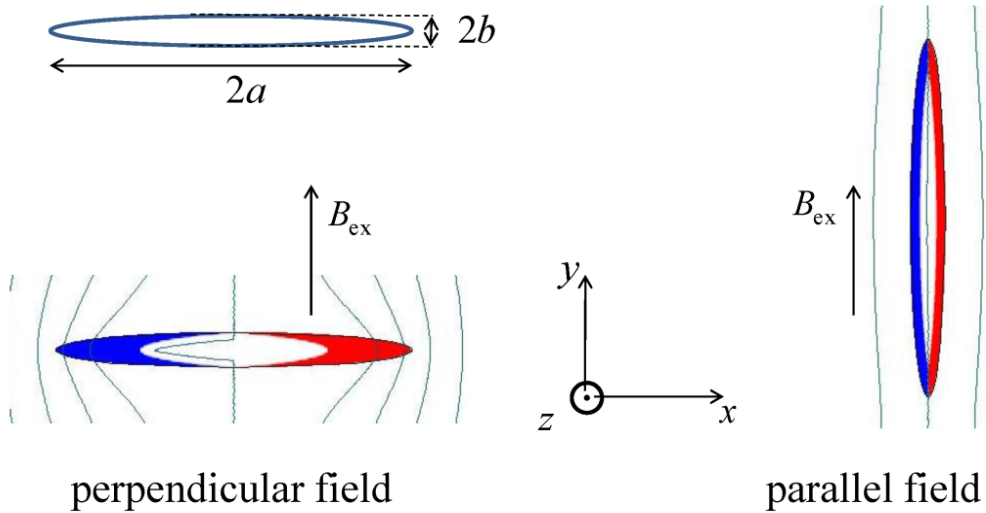

**Fig. 11:** Exposure of wire with elliptical cross-section to a perpendicular or parallel magnetic field

Let us now analyse the difference in the magnetic flux dynamics between two orientations of a wire with an elliptical cross-section. For this purpose we compare the magnetization loops calculated in two cases (by a numerical procedure) and plotted in Fig. 12. The difference in the loop 'thickness' controlled by the saturation magnetization is spectacular. This is because the same ratio ($a/b = 10$) exists for the values of $M_s$. Much less prominent is the difference in the penetration fields, $B_p$, which can be identified as the field occurring when the magnetization reaches its saturation value. The flux lines before entering the wire deform in order to adapt to become roughly parallel to the surface. This buckling creates more line tension in the perpendicular orientation; therefore the push to enter the wire is bigger, and the penetration process is quicker than in the parallel case. However, this difference is far lower (a factor of about 2 in this case) than in the saturation magnetization. Because the dissipation of the electromagnetic energy in one cycle—commonly called the AC loss—per unit length of wire is given by a formula indicating that it is proportional to the loop area,

$$Q_l = \oint m \, dB_{ex} \quad [\text{J·m}^{-1}], \tag{13}$$

the AC loss in the perpendicular field is much bigger than in the parallel field.

The main consequences for electromagnets manufactured from wires containing hard superconductors, arising from the existence of magnetization currents in these materials, are as follows:

- magnetization currents generate a macroscopic magnetic field that is not proportional to the supplied current, e.g. after switching off the current a remanent field remains in the bore;
- ramping of the magnetic field generates a dissipation that warms up the magnet winding, so a cooling system must be designed to remove this AC loss.

In a magnet designed to use the transport capacity of a superconducting material in a reasonable way, the magnetization currents saturate the wires in a substantial portion of the winding. In other words, the local magnetic field is far larger than the penetration field of the wire, $B_p$. Then the essential parameter controlling the behaviour is the saturation magnetization. The volume density of the AC loss — i.e. the dissipation $Q$ released in the superconductor volume $V$ — in the cycle with AC field amplitude $B_a \gg B_p$ can be estimated as

$$\frac{Q}{V} = 4 M_s B_a . \tag{14}$$

Two cases for the orientation of the magnetic field with respect to a wire of elliptic cross-section with aspect ratio 10 will then result in the prediction of an AC loss 10 times higher in the perpendicular case because of the difference in the saturation magnetization.

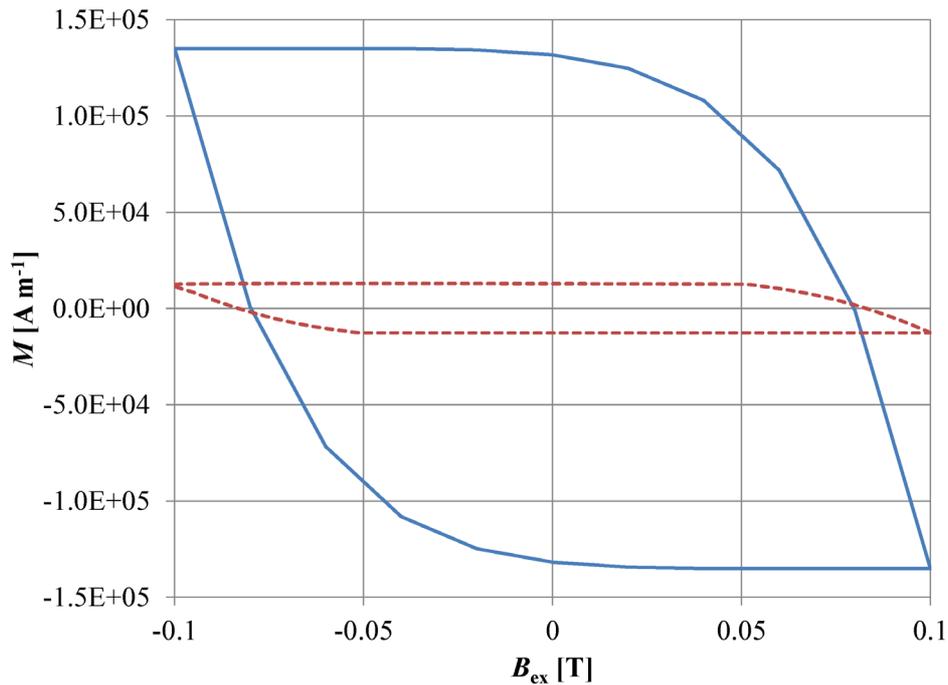

**Fig. 12:** Magnetization loops calculated for a wire of elliptical cross-section, where $a$ = 1 mm, $b$ = 0.1 mm, $I_c$ = 100 A, exposed to the magnetic fields of parallel (dashed) and perpendicular (solid) orientation with respect to the major semi-axis.

### 2.2.3 Slab in parallel field

The distributions shown in Figs. 8 and 11 and the magnetization loops plotted in Figs. 9 and 12 have been calculated numerically because there is no analytical solution available for the case of a wire magnetized in a perpendicular field. One can avoid setting up a numerical model by using the analytical approximation of these results [12]. On the other hand, analytical formulas are available for the case of a superconducting slab of thickness $w$, as shown in Fig. 1. Magnetic flux profiles and current distributions obtained at a cyclical change of magnetic field applied parallel to the slab are plotted in Fig. 13 for the field amplitude just reaching the penetration field value and in Fig. 14 for larger field amplitudes. The value of $B_p$ is obtained with the help of Eq. (2) under the assumption that the field just reaches the centre of the slab, which is at a distance $w/2$ from the surface:

$$B_{p,\text{slab}} = \mu_0 j_c \frac{w}{2}. \tag{15}$$

The saturation magnetization is easily calculated from Eq. (10) by taking into account that, for a slab with thickness $w$ in a parallel applied field, $\bar{x} = w/4$:

$$M_{s,\text{slab}} = j_c \frac{w}{4} = \frac{B_{p,\text{slab}}}{\mu_0}. \tag{16}$$

The volume density of loss in cyclic magnetization in an AC field with amplitude $B_a$ was derived for this case [3] as follows:

$$\frac{Q}{V} = \frac{1}{\mu_0} \begin{cases} \dfrac{2}{3} \dfrac{B_a^3}{B_p} & \text{for } B_a < B_p, \\ 2B_p B_a - \dfrac{4}{3} B_p^2 & \text{for } B_a > B_p, \end{cases} \tag{17}$$

with $B_p = B_{p,\text{slab}}$ given by Eq. (15).

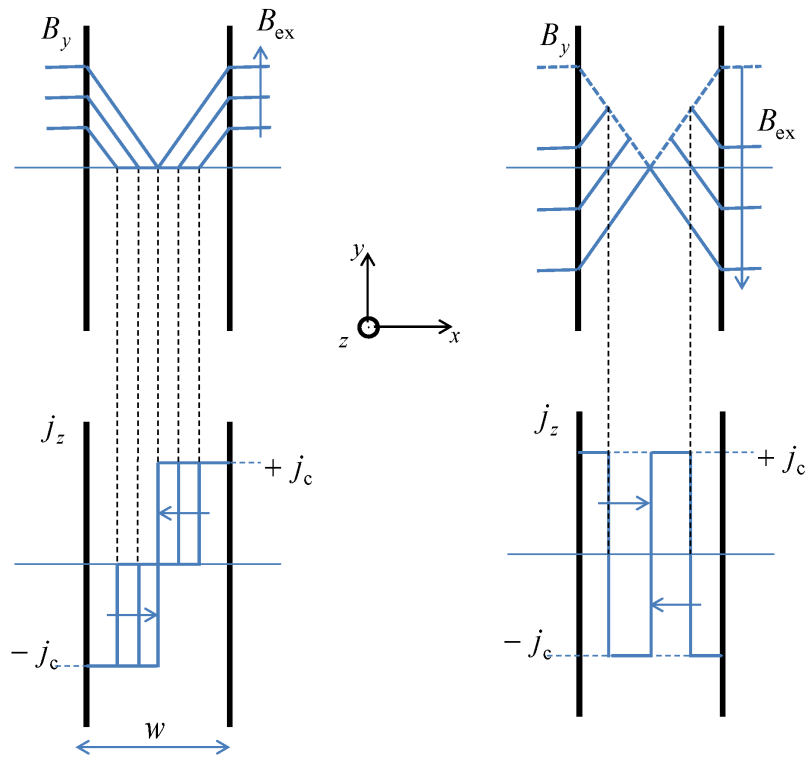

**Fig. 13:** Series of profiles of magnetic field (upper part) and electrical current density (lower part) at the first increase (left) and subsequent reduction (right) of the magnetic field applied in a parallel direction to a slab infinite in the $y$–$z$ plane. A particular case of maximal field equal to the penetration field, $B_p$, is shown.

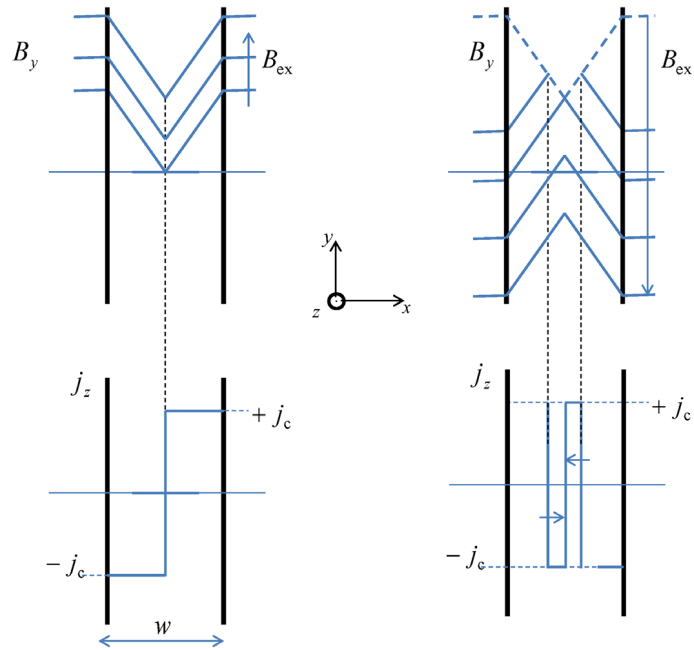

**Fig. 14:** Similar to Fig. 13, but the field increases from $B_p$ (left) and is then reduced to a negative value (right). Note that the entire section of the slab is filled with critical current density.

The plot of this dependence for slabs with different thicknesses and critical current densities is shown in Fig. 15. Among other things, it illustrates the fact that two different slabs exhibit the same volume density of loss provided the penetration field, $B_p$, is the same. An important feature of loss dependence is the change of slope when the amplitude trespasses the value of $B_p$: from $Q \propto B_a^3$ observed below penetration, it reduces to $Q \propto B_a$ for large amplitudes $B_a \gg B_p$ in accordance with the general formula (14). In spite of fact that the slab geometry is far from describing a wire used to wind a superconducting magnet, Eq. (17) was widely used to predict the loss in superconducting magnets developed for a generation of magnetic fields varying in time, such as dipoles for particle accelerators.

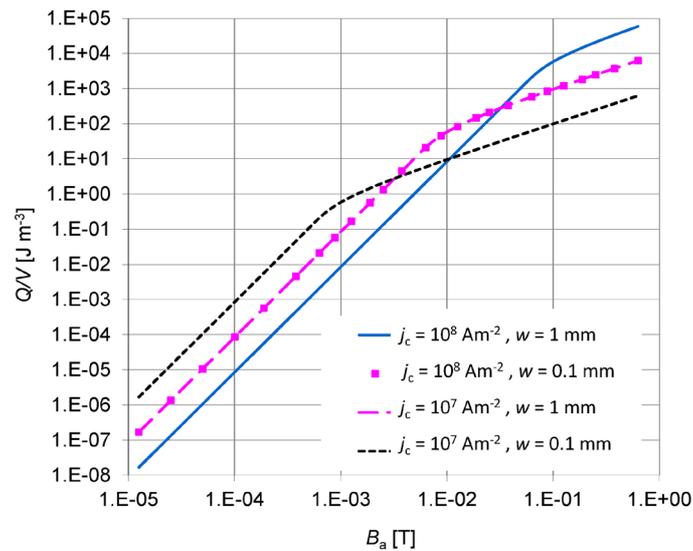

**Fig. 15:** Dependence of the volume density of AC loss calculated with the help of Eq. (17) for four slabs differing in thickness $w$ and critical current density $j_c$. The losses calculated for two slabs with identical values of the penetration field calculated from Eq. (15) are the same.

Nowadays, with increased computing power and numerical methods developed for the calculation of magnetic flux dynamics in hard superconductors, such approach serves only for qualitative considerations and rough estimations. Interestingly, the results [e.g. 13, 14] found by such computations always follow the general features shown in Fig. 16: there is a change in the slope at amplitudes comparable to the penetration field and an inverse order of loss level below and above this 'knee'. At low amplitudes, a smaller $B_p$ means higher losses; at large amplitudes, it results in lower losses.

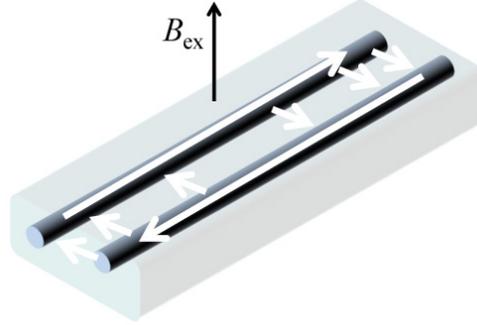

**Fig. 16:** Electrical loop created by the currents coupling two superconducting filaments when exposed to the magnetic field $B_{ex}$ changing in time (in this particular case, it is increasing).

## 3 Multicore wires and coupling currents

One important consequence of Eq. (10) is that the only way to reduce the magnetization currents without lowering the transport capacity of a superconducting wire—as well as the AC loss at large AC field amplitudes given by Eq. (14)—is to minimize the dimension of the wire that is perpendicular to the applied field. In the case of flat conductors, such as Rutherford cables or tapes coated with high-temperature superconductors, this means one must avoid the exposure of the magnetic field perpendicular to the flat face. This is not always possible, and therefore a remedy was developed that involved splitting the superconducting core into many filaments. Imagine dividing a wire of circular cross-section with radius $R_0$ into $N$ circular filaments. Maintaining the critical current requires that the total cross-section of the superconducting material be the same, so the filament radius will be $R_N = R_0/\sqrt{N}$. Then, according to Eq. (10), the saturation magnetization will be reduced by a factor $\sqrt{N}$. Therefore the wires used in magnets generating a pulse or AC magnetic field are composites containing fine (5–50 $\mu$m in diameter) superconducting filaments in a metallic matrix. The matrix—besides assisting thermal and mechanical stabilization—allows a transfer of electrical current in the transverse direction and the optimal distribution of this current among the superconducting filaments. However, another phenomenon influencing the macroscopic magnetization and the AC loss will appear at the ramp of the magnetic field: electrical currents connecting the filaments across the metallic matrix in closed loops, commonly called the coupling currents.

The mechanism of magnetically induced coupling is illustrated schematically for just two filaments in Fig. 16. The current loop provoked by the change in applied magnetic field consists of the part running along the superconducting filaments, but it must be closed either across the normal conducting matrix or at the ends of filaments. Let us now compare the flux dynamics in two perfectly coupled parallel round filaments for the case when the filaments would be completely insulated. Distributions of current density and magnetic field calculated for these two cases are shown for some representative instants of the magnetization cycle in Fig. 17. When the connection between filaments is perfect, all the current running along one filament will return through the second one:

$$\int_{1,coupled} j\,dS = -\int_{2,coupled} j\,dS, \qquad (18)$$

where the indices 1 and 2 denote the cross-sections of the two filaments, respectively. In the case of non-existing coupling,

$$\int_{1,\text{uncoupled}} j \, dS = \int_{2,\text{uncoupled}} j \, dS = 0, \quad (19)$$

i.e. all the current induced by the field change must return through the same filament. Let us evaluate the transverse dimension in these two cases. For uncoupled round filaments, each with radius $R$, the value of $\bar{x}$ is the same as for the single round wire with radius $R$:

$$\bar{x}_{2,\text{uncoupled}} = \frac{4R}{3\pi}. \quad (20)$$

In the case of perfect coupling, the maximum magnetic moment is determined by the distance between filaments, $d$, and one can estimate that

$$\bar{x}_{2,\text{coupled}} = \frac{d}{2}. \quad (21)$$

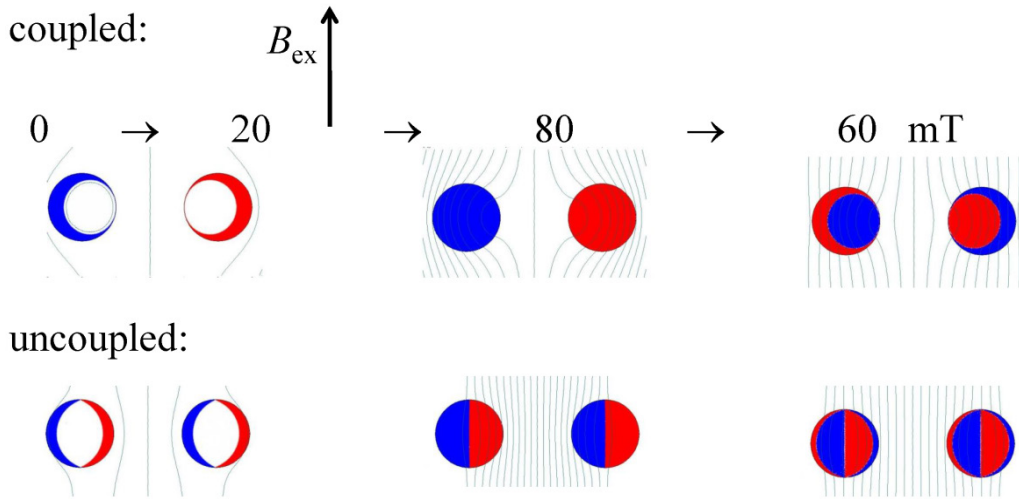

**Fig. 17:** Distributions of magnetic field (lines of constant vector potential) and electrical current density (grey scale) calculated for a pair of round wires during the change in applied magnetic field, $B_{\text{ex}}$ in a direction transverse to the plane connecting the axis of the two wires. The upper part shows the result obtained when assuming a perfect galvanic connection at the terminations of the wires; the lower part shows the distributions calculated for two insulated wires.

Magnetization loops calculated by a numerical finite element method [10] are given for these two cases in Fig. 18. The proportion of values for the saturation magnetization is given by the ratio of $\bar{x}$, which is evaluated using Eqs. (20) and (21). Using the values $R = 1$ mm and $d = 4$ mm, one finds that $M_s$ in the coupled case should be 4.7 times higher than in the uncoupled case, which is in good agreement with the numerically calculated result.

The existence of coupling currents means that splitting a single superconducting core into many filaments is not in itself sufficient to depress the magnetization currents and reduce AC loss. An additional measure is required to uncouple the filaments. This was achieved by means of a transposition through twisting the whole filamentary zone. The polarity of the electrical field induced by the change in the external magnetic field alternates in the half-loops between filaments, as shown in Fig. 19. The net voltage generated along one filament is therefore negligible, and thus there is no driving force to create interfilament currents. However, within one half of the twist pitch, $l_p$, a potential difference between parallel filaments remains, leading to a current traversing the matrix. This mechanism can be interpreted as a diffusion of magnetic flux opposed by coupling currents [1] with time constant

$$\tau = \frac{\mu_0}{2\rho_t}\left(\frac{l_p}{2\pi}\right)^2, \qquad (22)$$

where $\rho_t$ is the effective transverse resistivity of the multifilament composite. This quantity is estimated from the resistivity of the matrix, $\rho_m$, and the volume fraction occupied in the wire by the superconductor, $\lambda$, taking into account that superconducting filaments provide shorts for currents:

$$\rho_t = \rho_m \frac{1-\lambda}{1+\lambda}. \qquad (23)$$

This is the lower limit of the effective transverse resistivity. For cases when the interfaces between the filaments and matrix create obstacles for the current flow, e.g. because the formation of an oxide layer, the values of transverse resistivity could be significantly higher.

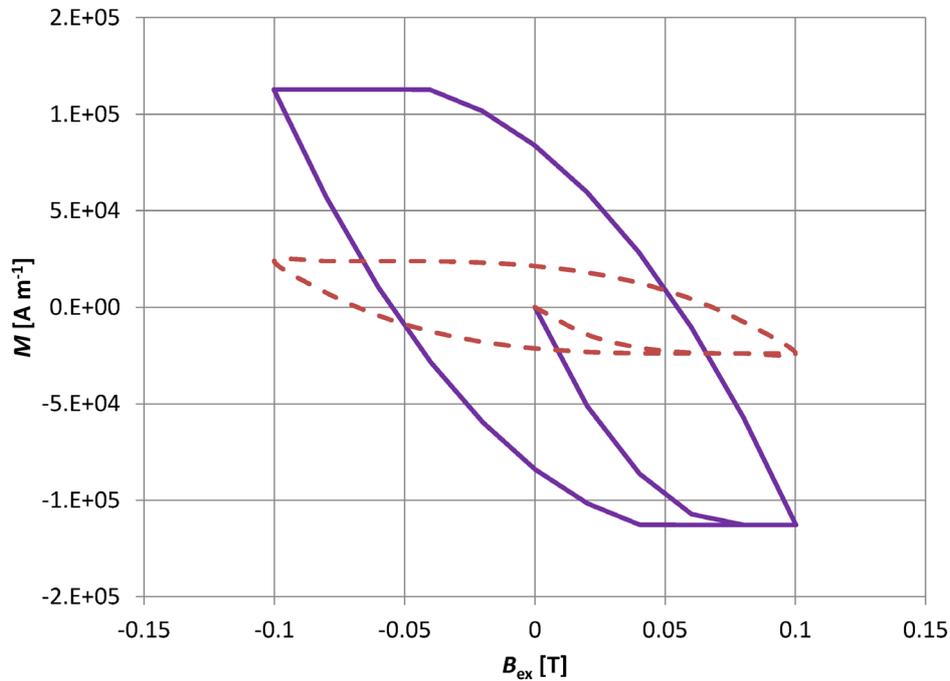

**Fig. 18:** Magnetization loops evaluated from the distributions shown in Fig. 17. Full line: coupled wires; dashed line: uncoupled wires.

In contrast to the magnetization currents generated by the flux pinning in a superconductor, the coupling currents represent a ramp-rate-dependent phenomenon. The screening field created by the coupling currents changes in time, with the characteristic time constant given by Eq. (22). This can be measured e.g. as the delay of the magnetic field in the closed vicinity of the wire with respect to the applied field, as illustrated in Fig. 20. Then the magnetization loops will be ramp-rate dependent and one must always indicate the waveform of the magnet current used in experiment, otherwise the interpretation of results is not possible.

Because the coupling currents are controlled by the normal resistance of the matrix, they exhibit properties similar to normal eddy currents. In particular, the behaviour at any cyclic change can be predicted on knowing the response at various frequencies. The volume density of the AC loss due to coupling currents caused by the external magnetic field $B_{ex} = B_a \sin(\omega t)$ is predicted by the following formula [15]:

$$\frac{Q_c}{V_w} = \frac{B_a^2}{\mu_0} \frac{\chi_0 \pi \omega \tau}{1+\omega \tau}, \qquad (24)$$

where $V_w$ is the volume of the whole wire (both superconductor and matrix) and $\chi_0$ is the magnetic susceptibility for a completely screened wire [16]. The latter quantity could be determined experimentally at low temperature and small $B_a$, or calculated from the shape of the wire. For the wires with cross-sections that could be approximated by an ellipse with axes $a$ and $b$, respectively, placed in a magnetic field oriented in parallel to the minor semi-axis, $b$, its value approaches $\chi_0 = 1 + a/b$ [17]. Accordingly, for the round wire, $\chi_0 = 2$. In weak magnetic fields with amplitudes well below the penetration field, the coupling loss should be corrected by this factor, taking into account a total expulsion of flux from the superconducting filaments; Eq. (24) is modified to

$$\left.\frac{Q_c}{V_w}\right|_{B_a \ll B_p} = \frac{B_a^2}{\mu_0}(1-\lambda)\frac{\chi_0 \pi \omega \tau}{1+\omega \tau}. \qquad (25)$$

In reality one often obtains a result that lies between the predictions of Eqs. (24) and (25), as shown in Fig. 21. In this plot the AC loss is expressed in terms of the imaginary part of the complex magnetic susceptibility [18],

$$\chi'' = \frac{Q}{V_w}\frac{\mu_0}{\pi B_a^2}. \qquad (26)$$

Such a representation is similar to the 'loss function' but allows better quantitative comparison.

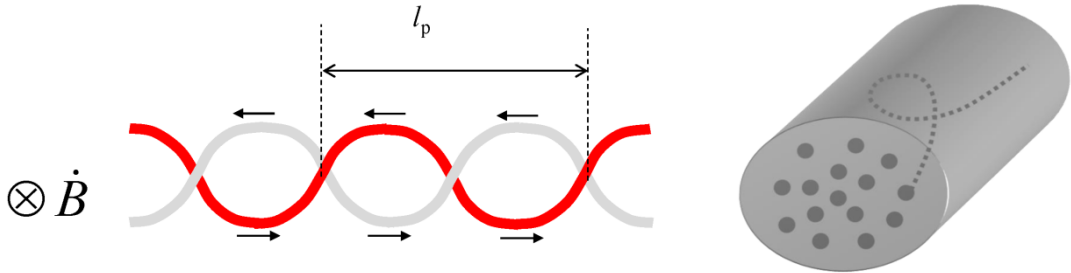

**Fig. 19:** The polarity of the electrical field induced between two twisted wires alternates every half-pitch

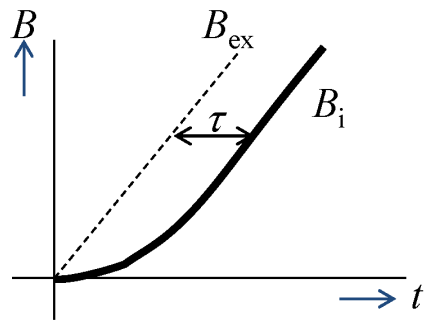

**Fig. 20:** The magnetic field inside or in the vicinity of the wire, $B_i$, compared to the applied field, $B_{ex}$, increasing with constant ramp rate, is delayed in time. From this delay, the time constant of the coupling currents, $\tau$, can be estimated.

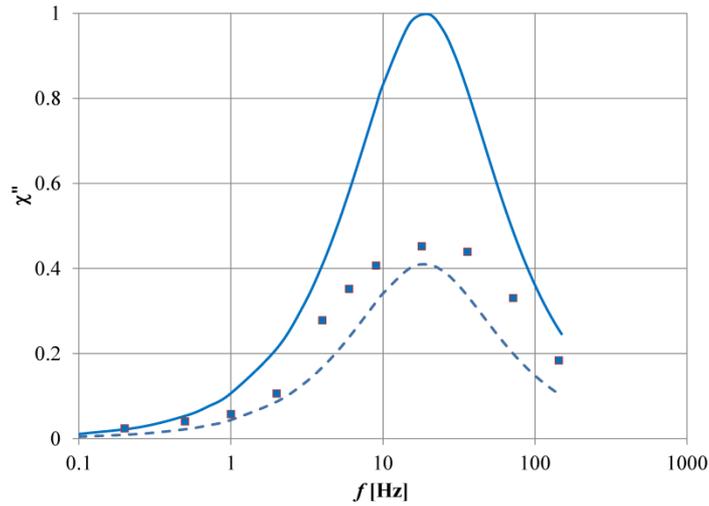

**Fig. 21:** Imaginary part of the complex magnetic susceptibility measured on a multifilamentary wire compared with two predictions that differ in the extent of filament penetration by an applied magnetic field. The solid line has been calculated from Eq. (24), assuming complete filament penetration; the dashed line is from Eq. (25), derived for completely screened filaments.

# 4    Conclusions

The pinning of magnetic flux, which is necessary to secure an exceptionally high current transport capacity in type II superconductors, is at the same time an obstacle when a variation in the magnetic field occurs. This is because a dissipation in the cyclic regime (e.g. when transporting AC current) appears. Also, a superconducting wire in a DC magnet must undergo a change in magnetic field at the ramp necessary to reach the operating field.

The dynamics of magnetic flux penetration into a hard superconductor (i.e. a type II superconductor able to pin the magnetic flux in its volume) is described by a complex process that depends on the properties of the superconducting material as well as the architecture of the superconducting wire and the orientation of magnetic field. Many basic features of this process can be modelled on a macroscopic scale with the help of the critical state model, assuming that, in a hard superconductor, the local value of the current density could be either zero or $j_c$.

Infiltration of a magnetic field provokes magnetization currents, which have negative consequences on the quality of the magnetic field generated by the superconducting magnet. These currents also lead to a dissipation of electromagnetic energy. The most practical way of minimizing these currents is through a reduction of the superconductor's dimension transverse to the magnetic field. This is why composite wires for pulsed magnets contain fine superconducting filaments, which in turn should be twisted in order to reduce the coupling currents induced in the interfilament loops. The basic principles erquired to understand these problems have been explained in this paper.

The current state of numerical methods allows detailed calculations that assume a more realistic description of the superconductor properties than is possible with the original critical state model with $j_c$ = constant. For example, a magnetic-field dependence of the critical current density and its anisotropy, and/or non-uniformity of $j_c$ in the volume of the superconductor, can be included in order to interpret the experimental data obtained from industrially produced materials [19]. There is currently a great deal of activity directed towards a more exact prediction of flux dynamics in non-traditional superconductors, and the topic is far from being completely managed. A recent review of the achievements in this field is given in [20].

## Appendix A

The quantity used here to define the magnetic fields is the induction, *B*, in units of tesla (T), where 1 T = 1 V·s·m$^{-2}$. This is the quantity measured by magnetic field sensors and an exerting force on a moving charged particle. In a material with a magnetic response, one finds loops of electrical currents (see Fig. A1, left panel). The area of the loop, $\vec{S}$, and the circulating current, *I*, determine its magnetic moment

$$\vec{m} = I\vec{S} \quad [\text{Am}^2]. \tag{A.1}$$

The magnetization of a sample with volume *V* is defined as the volume density of magnetic moments

$$\vec{M} = \frac{\sum \vec{m}}{V} \quad [\text{A·m}^{-1}]. \tag{A.2}$$

Quite often in the literature on superconductors an alternative definition is used that expresses the magnetization in tesla. This is obtained by multiplying Eq. (A.2) by the permeability of vacuum, $\mu_0 = 4\pi \cdot 10^{-7}$ H·m$^{-1}$. In all other aspects they are identical.

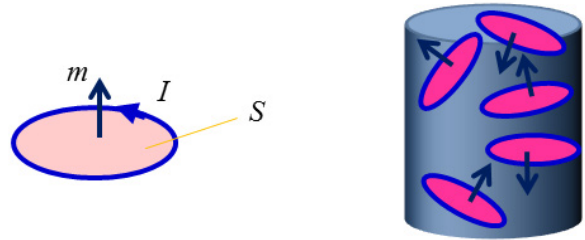

**Fig. A1:** Magnetic moment of a circular loop (left) and the magnetization of a macroscopic sample (right)

## Acknowledgement

This work was supported in part by the VEGA grant agency under contract No. 1/0162/11, by the European Commission EURATOM project FU07-CT-2007-00051 and co-funded by the Slovak Research and Development Agency under the contract No. DO7RP-0003-12.